\documentclass[aps,prd,amsmath,two column,nofootinbib,amssymb,showpacs]{revtex4}
\usepackage{amssymb}
\usepackage{txfonts}
\usepackage{amsfonts}
\usepackage{mathrsfs}
\usepackage{epsfig,bm,dcolumn}
\usepackage{graphicx}
\usepackage{color}
\usepackage{amsmath}
\usepackage{dcolumn}
\usepackage{overpic}
\usepackage{slashed}

\begin{document}
\title{Chiral phase transition and Schwinger mechanism in a pure electric field}
\author{Gaoqing Cao and Xu-Guang Huang}
\affiliation{Physics Department and Center for Particle Physics and Field Theory, Fudan University, Shanghai 200433, China.}

\date{\today}
\begin{abstract}
We systematically study the chiral symmetry breaking and restoration in the presence of a pure electric field in the Nambu--Jona-Lasinio (NJL) model at finite temperature and baryon chemical potential. In addition, we also study the effect of the chiral phase transition on the charged pair production due to the Schwinger mechanism. For these purposes, a general formalism for parallel electric and magnetic fields is developed at finite temperature and chemical potential for the first time. In the pure electric field limit $B\rightarrow0$, we compute the order parameter, the transverse-to-longitudinal ratio of the Goldstone mode velocities, and the Schwinger pair production rate as functions of the electric field. The inverse catalysis effect of the electric field to chiral symmetry breaking is recovered. And the Goldstone mode is find to disperse anisotropically such that the transverse velocity is always smaller than the longitudinal one, especially at nonzero temperature and baryon chemical potential. As expected, the quark-pair production rate is greatly enhanced by the chiral symmetry restoration.
\end{abstract}
\pacs{11.30.Qc, 05.30.Fk, 11.30.Hv, 12.20.Ds}

\maketitle
\section{Introduction}
Chiral symmetry and its spontaneous breaking is of fundamental importance for the quantum chromodynamics (QCD) as it explains the dynamical origin of the masses of hadrons. About two decades ago, it was revealed that the presence of a magnetic field would enhance the chiral condensate at zero temperature and zero quark chemical potentials~\cite{Klevansky:1989vi,Suganuma:1990nn,Klimenko:1991he,Klimenko:1992ch,Krive:1992xh} --- a phenomenon later known as the magnetic catalysis of chiral symmetry breaking (CSB)~\cite{Gusynin:1994re,Gusynin:1994xp,Gusynin:1995nb}, see recent review~\cite{Miransky:2015ava}. Quite recently, the lattice QCD simulations showed that in the temperature region near the critical temperature $T_c$ of the chiral phase transition, the effect of the magnetic field on the CSB is very different from that at zero temperature: the presence of the magnetic field tends to restore rather than break the chiral symmetry~\cite{Bali:2011qj,Bali:2012zg,Bruckmann:2013oba}. This inverse magnetic catalysis of CSB near $T_c$ seems very surprising and attracts a lot of theoretical interests, but it is still not fully understood, see e.g. Refs.~\cite{Fukushima:2012xw,Fukushima:2012kc,Kojo:2012js,Chao:2013qpa,Yu:2014sla,Feng:2014bpa,Cao:2014uva,Ferrer:2014qka,Farias:2014eca,Mueller:2015fka}. At zero temperature but finite quark chemical potential, analogous inverse magnetic catalysis was also found~\cite{Preis:2010cq}.

Where can strong magnetic fields be generated? In nature, the neutron stars especially the magnetars may have surface magnetic fields of the order $10^{14}-10^{15}$ Gauss~\cite{Olausen:2013bpa,Turolla:2015mwa}. In experiments, recently, it was revealed that very strong magnetic fields can be generated in high-energy peripheral heavy-ion collisions~(HICs)~\cite{Skokov:2009qp,Voronyuk:2011jd,Bzdak:2011yy,Deng:2012pc}: the numerical studies showed that the magnetic field in RHIC Au + Au collisions at $\sqrt{s}=200$ GeV can reach $5-6 m_\pi^2$ while in LHC Pb + Pb collisions at $\sqrt{s}=2.76$ TeV can reach $70 m_\pi^2$ where $m_\pi$ is the pion mass. These strong magnetic fields in HICs may drive the charge separation with respect to the reaction plane and the splitting of the elliptic flows of the charged pions through the underlying chiral magnetic and separation effects~\cite{Kharzeev:2007jp,Fukushima:2008xe,Son:2004tq,Metlitski:2005pr,Burnier:2011bf}, see Refs.~\cite{Kharzeev:2013ffa,Kharzeev:2015kna,Liao:2014ava,Huang:2015oca} for review.

In the HICs, the electric fileds can also be generated owing to the event-by-event fluctuations~\cite{Bzdak:2011yy,Deng:2012pc,Bloczynski:2012en,Bloczynski:2013mca} or in asymmetric collisions like Cu + Au collision~\cite{Hirono:2012rt,Deng:2014uja,Voronyuk:2014rna}, and the strength of the electric fields can be roughly of the same order as the magnetic fields. These strong electric fields can lead to anomalous transport phenomena in HICs as well, that is, chiral electric separation effect~\cite{Huang:2013iia,Jiang:2014ura,Pu:2014cwa,Ma:2015isa} and other novel observations like the charge dependence of the directed flow in Cu + Au collisions~\cite{Hirono:2012rt,Voronyuk:2014rna}.

The strong electric fields in HICs naturally inspire us to consider the effect of the electric field on the chiral phase transition. In this paper, we will systematically study the effect of a pure electric field on the chiral symmetry breaking and restoration in the framework of the Nambu--Jona-Lasinio (NJL) model. In fact, the effect of electric field had been previously studied at zero temperature many years ago and it was discovered that the electric field always tends to restore the chiral symmetry~\cite{Klevansky:1989vi}. The underlying mechanism is simple: the electric field always tends to break the quark-antiquark pair constituting the chiral condensate which triggers the chiral symmetry breaking. Another work that concerned about the effect of the second Lorentz invariant $I_2=\bf{B\cdot E}$ also found that the presence of the electric field in $I_2$ would suppress the chiral condensate at zero temperature~\cite{Babansky:1997zh}. More recently, a detailed study of the stability of a chiral symmetry breaking system in electric field was performed in the framework of the chiral perturbation theory~\cite{Cohen:2007bt}. As had been illuminated in Schwinger's seminal work, the electric field would induce pair production and the production rate is closely related to the relative magnitude of field strength and the charged particle mass~\cite{Schwinger:1951nm}. The Schwinger mechanism has been explored in different physical contexts, see e.g., Refs.~\cite{Kim:2000un,Nayak:2005pf,Gavrilov:2007hq,Gavrilov:2006jb,Yamamoto:2012bd}, but as far as we know, none of the previous works has combined the chiral phase transition with the pair production mechanism in a single quark model. In our opinion, the presence of the electric field will on one hand modify the QCD vacuum (i.e. suppress the chiral condensate) and on the other hand create quark-antiquark pairs on top of the modified vacuum. Thus, it will be interesting to study how this twofold effect of the electric field works in detail. In addition, when the temperature and the quark chemical potential are finite, richer phenomena are expected to emerge and these, to our best knowledge, have not been addressed so far.

The paper is organized as follows. In Sec.\ref{sec1}, we establish a general formalism for systems with parallel electric and magnetic fields at finite temperature and baryon chemical potential. Due to the non-renormalizability of NJL model, a proper regularization scheme is introduced to both the gap equation and the expansion coefficients in Sec.\ref{regularization}. In Sec.\ref{sec2}, we present our numerical calculations for the cases with vanishing temperature and finite temperature, respectively. Finally, a summary is given in Sec.\ref{sec3}.

\section{Formalism}\label{sec1}
\subsection{NJL model in the case $\bf{B\parallel E}$}
In order to study the chiral symmetry breaking and restoration of quark matter, we adopt the NJL model which has the same approximate chiral symmetry as QCD. We will consider a background with constant parallel electric and magnetic fields and with a baryon chemical potential $\mu$. It is more convenient to express the Lagrangian density in Euclidean space,
\begin{eqnarray}
{\cal L}=\bar\psi\left(i\slashed{D}-m_0-i{\mu}\gamma^4\right)\psi
+{G}\left[\left(\bar\psi\psi\right)^2+\left(\bar\psi i\gamma_5{\boldsymbol \tau}\psi\right)^2\right],
\end{eqnarray}
where $\psi=\left(u,d\right)^T$ is the two-flavor quark field, $m_0$ is the current quark mass, $G$ is the coupling constant with dimension GeV$^{-2}$ and ${\boldsymbol \tau}$ are pauli matrices in flavor space. Here, $\slashed{D}=\sum_{\mu=1}^4(\partial_\mu-iQ A_\mu)\gamma_\mu$~($\gamma_i=\gamma^i$) is the covariant derivative with the electric charge matrix $Q={\rm diag}(2e/3,-e/3)$ in flavor space and the vector potential in Euclidean space chosen as $A_\mu=(iEz,0,-Bx,0)$ which stands for electric and magnetic fields both along $z$-direction without loss of generality. In order to study the ground state of the system, we introduce four auxiliary fields $\sigma=-2G\bar\psi\psi$ and ${\boldsymbol \pi}=-2G\bar\psi i\gamma_5{\boldsymbol \tau}\psi$, and the Lagrangian density becomes
\begin{eqnarray}
{\cal L} &=& \bar\psi\bigg[i{\slashed D}-m_0-\sigma-i\gamma_5\left(\tau_3\pi_0+\tau_\pm\pi_\pm\right)-i{\mu}\gamma^4\bigg]\psi\nonumber\\
&&-{\sigma^2+\pi_0^2+\pi_\mp\pi_\pm\over 4G},
\end{eqnarray}
where the physical iso-vector fields $\pi_\pm$ are related to the auxiliary fields as $\pi_\pm=\left(\pi_1\mp i\pi_2\right)/\sqrt 2$, and $\tau_\pm=\left(\tau_1\pm i\tau_2\right)/\sqrt 2$ are the raising and lowering operators in flavor space, respectively.

The presence of the electromagnetic field breaks the isospin symmetry explicitly, that is, $SU(2)\rightarrow U_3(1)$. The order parameters for the spontaneous chiral symmetry breaking and $\tau_3$-isospin symmetry breaking can be chosen respectively the expectation values of the physical fields $\langle\sigma+i\gamma_5\tau_3\pi_0\rangle$ and $\langle \pi_\pm\rangle$. We choose $\langle\pi_0\rangle=0$ for simplicity~\footnote{In a previous study, we showed that once $I_2\neq0$ the QCD vacuum will inevitably contain the component of neutral pion condensate. However, the main purpose of the present paper is to study the pure electric field case where no external source is about to induce the $\pi^0$ condensation~\cite{Cao:2015cka}. Thus we set $\langle\pi_0\rangle=0$ here after.} and $\langle\pi_\pm\rangle=0$ because we will consider vanishing isospin chemical potential. Then, by taking the relationship between the chiral condensate and the dynamical mass $\langle\sigma\rangle=m-m_0$ into account and integrating out the quark degrees of freedom, the partition function ${\cal Z} =\int[{\cal D}\psi][{\cal D}\bar\psi]e^{\int d^4x {\cal L}}$ can be expressed in a bosonic degree of freedom only,
\begin{eqnarray}
{\cal Z}\!\!&=&\!\!\!\!\int\![{\cal D}\hat{\sigma}][{\cal D}\hat{\pi}_0][{\cal D}\hat{\pi}_\pm]\exp\Big\{\!-\!\!\int\!\!dx\big[ {\left(m-m_0\right)^2\!+\!\hat{\sigma}^2\!+\!\hat{\pi}_0^2\!+\!\hat{\pi}_\pm^2\over 4G}\big]\nonumber\\
&&+\text{Tr}\ln\left[i{\slashed D}-m-\hat{\sigma}-i\gamma_5\left(\tau_3\hat{\pi}_0+\tau_\pm\hat{\pi}_\pm\right)-i{\mu}\gamma^4\right]\Big\},
\end{eqnarray}
where the fields with hat denote the bosonic fluctuations and the trace is taken over the quark spin, flavor, color, and the space-time coordinate spaces. In mean field approximation, the thermodynamic potential can be directly obtained as
\begin{eqnarray}
\Omega(m)={(m-m_0)^2\over 4G}-{1\over\beta V}\text {Tr}\ln\left[i{\slashed D}-m-i{\mu}\gamma^4\right],
\end{eqnarray}
where $\beta=1/T$ is the inverse temperature and $V$ is the volume of the system. Then, the gap equation can be formally derived by the extremal condition $\partial\Omega/\partial m=0$,
\begin{eqnarray}\label{gap1}
{m-m_0\over 2G}-{1\over\beta V}\text {Tr}{\cal S}(x,x')=0,
\end{eqnarray}
where ${\cal S}(x,x')\!=\!-(i{\slashed D}-m-i{\mu}\gamma^4)^{-1}\delta(x-x')$ is the fermion propagator which is consistent with that defined in Schwinger's work~\cite{Schwinger:1951nm}.

The bosonic fluctuations also contribute to the thermodynamic potential and their propagators satisfy
\begin{eqnarray}
i{\cal D}_M^{-1}(x,x')&=&\frac{\delta(x-x')}{2G}+\Pi_M(x,x')\nonumber\\
&=&\!\!\!\!\frac{e^{iq_{M}\!\!\int_{x'}^x\!A dx}}{2G}\delta(x\!-\!x')\!+\!\text{Tr}{\cal S}(x,\!x')\Gamma_M{\cal S}(x',\!x)\Gamma_M^*,
\end{eqnarray}
where $\Pi_M(x)$ are polarization functions and the interaction vertices are given by
\begin{eqnarray}
\begin{array}{cc}
\Gamma_M=\left\{\begin{array}{ll}
1,&M=\hat\sigma\\
i\gamma_5\tau_+,&M=\hat{\pi}_+\\
i\gamma_5\tau_-,&M=\hat{\pi}_-\\
i\gamma_5\tau_3,&M=\hat{\pi}_0\\
\end{array}\right. ;
&
\Gamma_M^*=\left\{\begin{array}{ll}
1,&M=\hat\sigma\\
i\gamma_5\tau_-,&M=\hat{\pi}_+\\
i\gamma_5\tau_+,&M=\hat{\pi}_-\\
i\gamma_5\tau_3,&M=\hat{\pi}_0\\
\end{array}\right. .
\end{array}
\end{eqnarray}
It is easy to verify that there is no mixing between different collective modes because of the absence of pion condensate.
For neutral modes, that is, $\hat\sigma$ and $\hat{\pi}_0$, the total Wilson lines vanish; thus, the propagators only depend on the relative variables $x-x'$ and can be transformed to energy-momentum space. Then, the dispersions of the collective modes can be obtained from the poles of the propagators,
\begin{eqnarray}\label{boson1}
i\tilde{\cal D}_M^{-1}(q)=\!\frac{1}{2G}\!+\!\tilde{\Pi}_M(q)\!=\!\frac{1}{2G}\!+\!\text{Tr}\tilde{\cal S}(p+q)\Gamma_i\tilde{\cal S}(p)\Gamma_i^*=0,
\end{eqnarray}
where $\tilde{\cal S}(p)$ is the effective fermion propagator in energy-momentum space which will be given later with Schwinger's approach and the trace is now taken over the quark spins, flavors, colors, and energy-momenta. It is easy to verify that in the chiral limit $m_0=0$, $\hat{\pi}_0$ is the Goldstone mode for chiral symmetry breaking even with the presence of the electromagnetic field. However, the charged modes $\hat{\pi}_\pm$ are no longer the Goldstone modes and we will not focus on them in the present study.

\subsection{Explicit expressions with Schwinger's approach}
In order to obtain an explicit formulism for further calculations, the most important mission is to find the explicit forms of quark propagators. In a constant electromagnetic field, the quark propagators can be given explicitly by using the Schwinger's approach~\cite{Schwinger:1951nm} as
\begin{widetext}
\begin{eqnarray}
{\cal S}_{\rm f}(x,x')&=&{-i\over(4\pi)^2}\int_0^\infty {ds\over s^2}\;e^{iq_{\rm f}\int_{x'}^x\tilde{A} dx}\big[-{1\over2}\gamma\big(q_{\rm f}F\coth(q_{\rm f}Fs)+q_{\rm f}F\big)(x-x')+m\big]\nonumber\\
&&\times\exp\Big\{-im^2s-L(s)+{i\over4}(x-x')q_{\rm f}F\coth(q_{\rm f}Fs)(x-x')+{i\over2}q_{\rm f}\sigma Fs\Big\}\nonumber\\
&=&{-i\over(4\pi)^2}\int_0^\infty {ds\over s^2}\;e^{iq_{\rm f}\int_{x'}^xAdx+(x_4-x'_4)\mu}\big[-{1\over2}\gamma\big(q_{\rm f}F\coth(q_{\rm f}Fs)+q_{\rm f}F\big)(x-x')+m\big]\nonumber\\
&&\times\exp\Big\{-im^2s+{i\over4}(x-x')q_{\rm f}F\coth(q_{\rm f}Fs)(x-x')+{i\over2}q_{\rm f}\sigma Fs\Big\}{-(q_{\rm f}s)^2I_2\over\text{Im}\cosh\big(iq_{\rm f}s(I_1+2iI_2)^{1/2}\big)},
\end{eqnarray}
\end{widetext}
where the baryon chemical potential is introduced by considering the effective vector potential
 $\tilde{A}_\mu\!=\!{A_\mu}+(-i\mu/q_{\rm f},0,0,0)$, $L(s)\!=\!{1\over2}\text{tr}\ln\big[(q_{\rm f}Fs)^{-1}\sinh(q_{\rm f}Fs)\big]$, $\sigma F=\sigma_{\mu\nu}F_{\mu\nu}$ with $\sigma_{\mu\nu}={i\over2}[\gamma_\mu,\gamma_\nu]$ the $4\times4$ tensor matrices, and $I_1\!=\!{\bf B}^2-{\bf E}^2$ and $I_2\!=\!{\bf B\cdot E}$ are the two Lorentz invariants for electromagnetic field. It is important to point out that the chemical potential is a prior constrained by certain periodicity condition at finite temperature and thus cannot be canceled out beforehand by the gauge transformation. The integration involved in the Wilson line is chosen along a straight line here and we will neglect it in the following discussion as it will not affect either the gap equations or the dispersions of neutral collective modes. Then, we can define the effective propagators of quarks which only depends on the relative displacement $x-x'$:
\begin{widetext}
\begin{eqnarray}
\tilde{\cal S}_{\rm f}(x-x')&=&{-i\over(4\pi)^2}\int_0^\infty {ds\over s^2}e^{(x_4-x'_4)\mu}{-(q_{\rm f}s)^2I_2\over\text{Im}\cosh\big(iq_{\rm f}s(I_1+2iI_2)^{1/2}\big)}\big[-{1\over2}\gamma\big(q_{\rm f}F\coth(q_{\rm f}Fs)+q_{\rm f}F\big)(x-x')+m\big]\nonumber\\
&&\times\exp\Big\{-im^2s+{i\over4}(x-x')q_{\rm f}F\coth(q_{\rm f}Fs)(x-x')+{i\over2}q_{\rm f}\sigma Fs\Big\}.
\end{eqnarray}
\end{widetext}
As can be seen, the effective propagators are also gauge invariant. According to the finite temperature quantum theory, the effective propagators should be (anti-)periodic in imaginary-time. This will be automatically satisfied if we take the following Fourier transformation:
\begin{widetext}
\begin{eqnarray}\label{Sf}
\tilde{\cal S}_{\rm f}(\tilde{p})&=&\int{d^4(x-x')}\tilde{\cal S}_{\rm f}(x-x')e^{-ip(x-x')}\nonumber\\
&=&-i\int_0^\infty {ds\over s^2}\det\big[i{q_{\rm f}F\coth(q_{\rm f}Fs)}\big]^{-1/2}{-(q_{\rm f}s)^2I_2\over\text{Im}\cosh\big(iq_{\rm f}s(I_1+2iI_2)^{1/2}\big)}\big[-\gamma\Big(1+{q_{\rm f}F\over q_{\rm f}F\coth(q_{\rm f}Fs)}\Big)\tilde{p}+m\big]\nonumber\\
&&\times\exp\Big\{-im^2s-{i}\tilde{p}{1\over q_{\rm f}F\coth(q_{\rm f}Fs)}\tilde{p}+{i\over2}q_{\rm f}\sigma Fs\Big\},
\end{eqnarray}
\end{widetext}
where $\tilde{p}\!=\!(p_4+i\mu,{\bf{p}})\!=\!(\omega_n+i\mu,\bf{p})$ with the fermionic Matsubara frequency $\omega_n\!=\!(2n+1)\pi T~(n\!\in\! Z)$.
The reason why the integration over $x_4-x'_4\in(-\beta/2,\beta/2]$ can be completed is that the effective integral variables are now $\big[q_{\rm f}F\coth(q_{\rm f}Fs)\big]^{1/2}(x-x')$ which vary from $-\infty$ to $\infty$ near the most significant point $s\!=\!0$. Usually, the matrix $-i{q_{\rm f}F\coth(q_{\rm f}Fs)}$ can not be guaranteed to be positive definite under the transformation $s\!\rightarrow\!-is$, thus this expression is only formal. On the other hand, because of the introduction of the proper time $s$, the collective modes can only be adequately evaluated in energy-momentum space when $\mu\neq0$. Therefore, the integrations in energy-momentum space should also be done formally to arrive at correct results. To make the procedure mathematically operable, we define
\begin{eqnarray}
\label{formu}
\int_{-\infty}^{\infty}e^{-a x^2}dx&=&{|a|^{1/2}\over a^{1/2}}\int_{-\infty}^{\infty}e^{-|a| x^2}dx,\nonumber\\
T\sum_{n=-\infty}^{\infty}e^{-a \omega_n^2}&=&{|a|^{1/2}\over a^{1/2}}T\sum_{n=-\infty}^{\infty}e^{-|a| \omega_n^2},
\end{eqnarray}
for any real parameter $a$. In this way, we are able to recover the correct gap equations at $T=0$, which can also be evaluated directly in coordinate space and thus free from the non-positive-definite problem. Thus the integration in Eq. (\ref{Sf}) should be understood in the same sense as in Eq. (\ref{formu}).

We now evaluate the effective propagators of quarks, then the gap equations and the inverse propagators of neutral collective modes explicitly in the presence of parallel electric and magnetic fields. In Euclidean space, the electromagnetic field strength tensor $F$ can be easily evaluated to have the following anti-diagonal form:\\
\begin{eqnarray}
F=\left(
\begin{array}{cccc}
0&0&0&-iE\\
0&0&-B&0\\
0&B&0&0\\
iE&0&0&0
\end{array}\right).
\end{eqnarray}
It is easy to check that $F^{2n}~(n\in N)$ is diagonal, thus the variable transformation matrix can be evaluated as
\begin{widetext}
\begin{eqnarray}\label{cothF}
{q_{\rm f}F\coth(q_{\rm f}Fs)}&=&{1\over s}\Big[1+\sum_{n=1}^\infty{2^{2n}B_{2n}(q_{\rm f}Fs)^{2n}\over(2n)!}\Big]
=\left(
\begin{array}{cccc}
q_{\rm f}E\coth(q_{\rm f}Es)&0&0&0\\
0&q_{\rm f}B\cot(q_{\rm f}Bs)&0&0\\
0&0&q_{\rm f}B\cot(q_{\rm f}Bs)&0\\
0&0&0&q_{\rm f}E\coth(q_{\rm f}Es)
\end{array}\right),
\end{eqnarray}
where the Taylor expansion form of the matrix $\coth(q_{\rm f}Fs)$ has been used and $B_{2n}$ is the Bernoulli number. The diagonal form of this matrix means no mixing between different space-time components and would make the following reductions much easier. The exponential term $\exp({i\over2}q_{\rm f}\sigma Fs)$ can also be evaluated explicitly by utilizing the general property $({1\over2}\sigma F)^2=I_1-2i\gamma^5I_2$~\cite{Schwinger:1951nm}:
\begin{eqnarray}\label{tensor}
\exp({i\over2}q_{\rm f}\sigma Fs)&=&\sum_{t=\pm}\Big[\cosh\big(iq_{\rm f}s\sqrt{I_1+2itI_2}\big){1-t\gamma^5\over2}+{1\over2}\sigma F{\sinh\big(iq_{\rm f}s\sqrt{I_1+2itI_2}\big)\over\sqrt{I_1+2itI_2}}{1-t\gamma^5\over2}\big)\Big]\nonumber\\
&=&\!\!\!\cos(q_{\rm f}Bs)\cosh(q_{\rm f}Es)\!+\!i\sin(q_{\rm f}Bs)\sinh(q_{\rm f}Es)\gamma^5\!\!+\!\sin(q_{\rm f}Bs)\cosh(q_{\rm f}Es)\gamma^1\gamma^2\!\!+\!i\cos(q_{\rm f}Bs)\sinh(q_{\rm f}Es)\gamma^4\gamma^3.
\end{eqnarray}
Then the explicit forms of the effective propagators of quarks are given by
\begin{eqnarray}
\tilde{\cal S}_{\rm f}(\tilde{p})&=&i\int_0^\infty {ds}\exp\Big\{-im^2s-{i\tanh(q_{\rm f}Es)\over q_{\rm f}E}(\tilde{p}_4^2+p_3^2)-{i\tan(q_{\rm f}Bs)\over q_{\rm f}B}(p_1^2+p_2^2)\Big\}\Big[-\gamma\tilde{p}+m+{i\tanh(q_{\rm f}Es)}(\gamma^4p_3-\gamma^3\tilde{p}_4)\nonumber\\
&&+{\tan(q_{\rm f}Bs)}(\gamma^1p_2-\gamma^2p_1)\Big]\Big[1+i\tan(q_{\rm f}Bs)\tanh(q_{\rm f}Es)\gamma^5+\tan(q_{\rm f}Bs)\gamma^1\gamma^2+i\tanh(q_{\rm f}Es)\gamma^4\gamma^3\Big].
\end{eqnarray}
We find that the electric and magnetic fields couple with the coordinate indices $3,4$ and $1,2$, separately.
Furthermore, the effective propagators are actually invariant under the following combined transformations:
\begin{eqnarray}
E\leftrightarrow iB,\ \gamma^4(\gamma^3)\leftrightarrow\gamma^1(\gamma^2),\ \tilde{p}_4({p}_3)\leftrightarrow{p}_1({p}_2),
\end{eqnarray}
which shows the duality between the electric and magnetic fields.

Armed with the effective propagators of quarks, it is straightforward but a little tedious to deduce the inverse propagators of the neutral collective modes. We put the details to Appendix.\ref{nmodes} and give directly the final results here,
\begin{eqnarray}\label{propagator}
i\tilde{\cal D}_M^{-1}(q)
&=&\frac{1}{2G}-4N_{c}\sum_{\rm f=u,d}\sum_{\tilde{p}}\int_0^\infty {ds}\int_0^\infty {ds'}\exp\Big\{-im^2s-{if_{s}[(\tilde{p}_4+q_4)^2+(p_3+q_3)^2]\over q_{\rm f}E}-{ig_{s}[(p_1+q_1)^2+(p_2+q_2)^2]\over q_{\rm f}B}\nonumber\\
&&-im^2s'-{if_{s'}(\tilde{p}_4^2+p_3^2)\over q_{\rm f}E}-{ig_{s'}(p_1^2+p_2^2)\over q_{\rm f}B}\Big\}\Big\{\alpha_Mm^2\big(1-g_{s}g_{s'}\big)\big(1+f_{s}f_{s'}\big)-\big[(\tilde{p}_4+q_4)\tilde{p}_4+({p}_3+q_3){p}_3\big]
\nonumber\\
&&\times\big(1-f_{s}^2\big)\big(1-f_{s'}^2\big)\big(1-g_{s}g_{s'}\big)-\big[({p}_2+q_2){p}_2+({p}_1+q_1){p}_1\big]\big(1+g_{s}^2\big)
\big(1+g_{s'}^2\big)\big(1+f_{s}f_{s'}\big)\Big\},
\end{eqnarray}
where $\sum_{\tilde{p}}=T\sum_{n}\int{d^3\bf{p}\over(2\pi)^3}$ for simplicity, $f_{s}=\tanh(q_{\rm f}Es)$, $g_{s}=\tan(q_{\rm f}Bs)$, and $\alpha_M=1(-1)$ for $\sigma(\pi_0)$ mode. The expansion coefficients of $i\tilde{\cal D}_0^{-1}(q)$ around small momenta ${\bf q}$ at zero energy $q_4=0$ are:
\begin{eqnarray}\label{xi}
\xi_{\bot}&=&-{N_{c}\over4\pi^{2}}\sum_{\rm f=u,d}\int_0^\infty {ds}\int_0^\infty {ds'}e^{-im^2(s+s')}{q_{\rm f}B\over i(g_{s}+g_{s'})}
{q_{\rm f}E\over i(f_{s}+f_{s'})}\vartheta_3\left({\pi\over 2}+i{\mu\over 2T},e^{-|i{q_{\rm f}E\over 4(f_s+f_{s'})T^2}|}\right)\Big\{{ig_{s}g_{s'}\over
 q_{\rm f}B(g_{s}+g_{s'})}\nonumber\\
&&\times\Big[m^2f_{s'}(f_{s}+f_{s'})(1-g_{s}g_{s'})-iq_{\rm f}B\Big({g_{s'}-g_{s}\over g_{s}(g_{s}+g_{s'})}
-{g_{s'}\over (1-g_{s}g_{s'})}\Big)
(1-f_{s'}^2)(1+g_{s}^2)(1-g_{s}g_{s'})\Big]\nonumber\\
&&+{2g_{s}g_{s'}\over (g_{s}+g_{s'})^2}(1+g_{s}^2)(1+g_{s'}^2)(1+f_{s}f_{s'})\Big\},\\
\xi_{3}&=&-{N_{c}\over4\pi^{2}}\sum_{\rm f=u,d}\int_0^\infty {ds}\int_0^\infty {ds'}e^{-im^2(s+s')}{q_{\rm f}B\over i(g_{s}+g_{s'})}
{q_{\rm f}E\over i(f_{s}+f_{s'})}\vartheta_3\left({\pi\over 2}+i{\mu\over 2T},e^{-|i{q_{\rm f}E\over 4(f_s+f_{s'})T^2}|}\right)\Big\{{if_{s}f_{s'}\over
 q_{\rm f}E(f_{s}+f_{s'})}\Big[m^2f_{s'}(f_{s}+f_{s'})\nonumber\\
&&\times(1-g_{s}g_{s'})-{iq_{\rm f}B\over (g_{s}+g_{s'})}(1+g_{s}^2)(1+g_{s'}^2)(1+f_{s}f_{s'})+iq_{\rm f}B\Big({1\over g_{s}+g_{s'}}
+{g_{s'}\over 1-g_{s}g_{s'}}\Big)
(1-f_{s'}^2)(1+g_{s}^2)(1-g_{s}g_{s'})\Big]\nonumber\\
&&+{f_{s'}\over f_{s}+f_{s'}}(1-f_{s}^2)(1-f_{s'}^2)(1-g_{s}g_{s'})\Big\},
\end{eqnarray}
\end{widetext}
where $\xi_\bot$ and $\xi_3$ correspond to the transverse and longitudinal motions relative to the direction of electric field respectively and $\vartheta_3(z,q)$ is the third Jacobi theta function obtained by working out the summation over the Matsubara frequency. It is worth mentioning that the expansion coefficient around small $q_4$ can only be correctly obtained when we sum over $\tilde{p}_4$ firstly in technique at finite temperature. Thus, the expansion coefficient around small $q_4$ can not be effectively evaluated by taking Taylor expansion beforehand as for the momenta. In the chiral limit, $\sqrt{\xi_{\bot}}$ and $\sqrt{\xi_{3}}$ are proportional to sound velocities along the transverse and longitudinal directions. Thus, their relative ratio will reflect whether any direction is more favored to the other in the electromagnetic field and deserves to study.

The dynamical mass $m$ involved in the formulas should be determined by the gap equation which can be obtained by following Eq. (\ref{gap1}), that is,
\begin{widetext}
\begin{eqnarray}
{m-m_0\over 2G}&=&4imN_{c}\sum_{\rm f=u,d}\sum_{\tilde{p}}\int_0^\infty {ds}\exp\Big\{-im^2s-{i\tanh(q_{\rm f}Es)\over q_{\rm f}E}(\tilde{p}_4^2+p_3^2)-{i\tan(q_{\rm f}Bs)\over q_{\rm f}B}(p_1^2+p_2^2)\Big\}\nonumber\\
&=&-i{mN_{c}\over4\pi^2}\sum_{\rm f=u,d}\int_0^\infty {ds}e^{-im^2s}\vartheta_3\left({\pi\over 2}+i{\mu\over 2T},e^{-|{iq_{\rm f}E\over 4\tanh(q_{\rm f}Es)T^2}|}\right)
{q_{\rm f}B\over\tan(q_{\rm f}Bs)}{q_{\rm f}E\over\tanh(q_{\rm f}Es)},
\end{eqnarray}
\end{widetext}
where the property of gamma matrices: $\text{tr}[\gamma^{\mu_1}\cdots\gamma^{\mu_{2n+1}}]\!=\!0$ $(n\in N)$ has been used. This result is consistent with that given in Ref.~\cite{Gusynin:1994re,Gusynin:1994xp,Gusynin:1995nb,Klevansky:1989vi,Babansky:1997zh} in the pure magnetic field case and the pure electric field case respectively at zero temperature limit. It should be pointed out that there is no Lorentz invariance in the system now because the finite temperature breaks it; but gauge invariance is always guaranteed. In the physical vacuum, that is, $T=0,\mu=0$, the gap equation is reduced to
\begin{eqnarray}
{m\!-\!m_0\over 2mG}\!=\!-i{N_{c}I_2\over4\pi^2}\!\!\!\sum_{\rm f=u,d}q_{\rm f}^2\!\int_0^\infty\!\!\!\!\!\!{ds}e^{-im^2s}{{\rm Re}\cosh[iq_{\rm f}s\!\sqrt{I_1\!-\!2iI_2}]\over {\rm Im} \cosh[iq_{\rm f}s\!\sqrt{I_1\!-\!2iI_2}]},
\end{eqnarray}
which is shown to be explicitly gauge and Lorentz invariant as expected.

\subsection{Regularization in the pure electric field case}\label{regularization}
Although the above formalism is general, we will hereafter focus on the pure electric field case. For the convenience of the numerical calculations, we first take a Wick rotation of the proper time integral and then transform the variable $s$ to $-is$ for all formulas. On the other hand, due to the non-renormalizability of NJL model, a proper regularization scheme should be introduced to take care of the nonphysical divergence. The regularization scheme we will adopt is similar to that developed in Ref.~\cite{Cao:2014uva,Cao:2015xja} in which the Goldstone theorem is self-consistently guaranteed with a single cutoff. Then the regularized gap equation in the pure electric field limit becomes
\begin{widetext}
\begin{eqnarray}
{m-m_0\over 2G}&=&{N_c m^2\over\pi^2}\left[\Lambda\sqrt{1+{\Lambda^2\over m^2}}-m\ln\left({\Lambda\over m}
+\sqrt{1+{\Lambda^2\over m^2}}\right)\right]-{N_cm\over\pi^2}\sum_{s=\pm}\int_0^\infty p^2 dp {1\over E(p)}{2\over 1+e^{(E(p)+s\mu)/T}}
\nonumber\\&&+{mN_{c}\over4\pi^2}\sum_{\rm f=u,d}\int_0^\infty {ds\over s^2}e^{-m^2s}\Big[\vartheta_3\left({\pi\over 2}+i{\mu\over 2T},e^{-\Big|{q_{\rm f}E\over 4\tan(q_{\rm f}Es)T^2}\Big|}\right){{q_{\rm f}Es\over\tan(q_{\rm f}Es)}}
-\vartheta_3\left({\pi\over 2}+i{\mu\over 2T},e^{-{1\over 4sT^2}}\right)\Big],\label{Mgap}
\end{eqnarray}
\end{widetext}
where $E(p)=\sqrt{p^2+m^2}$. With this in hand, the explicit form of the thermodynamic potential can be obtained by integrating the gap equation over $m$, that is,
\begin{widetext}
\begin{eqnarray}
\Omega(m)&=&\int dm \Big\{{m-m_0\over 2G}-{N_c m^2\over\pi^2}\left[\Lambda\sqrt{1+{\Lambda^2\over m^2}}-m\ln\left({\Lambda\over m}
+\sqrt{1+{\Lambda^2\over m^2}}\right)\right]+{N_cm\over\pi^2}\sum_{s=\pm}\int_0^\infty p^2 dp {1\over E(p)}{2\over 1+e^{(E(p)+s\mu)/T}}
\nonumber\\&&-{mN_{c}\over4\pi^2}\sum_{\rm f=u,d}\int_0^\infty {ds\over s^2}e^{-m^2s}\Big[\vartheta_3\left({\pi\over 2}+i{\mu\over 2T},e^{-\Big|{q_{\rm f}E\over 4\tan(q_{\rm f}Es)T^2}\Big|}\right){{q_{\rm f}Es\over\tan(q_{\rm f}Es)}}
-\vartheta_3\left({\pi\over 2}+i{\mu\over 2T},e^{-{1\over 4sT^2}}\right)\Big]\Big\}\nonumber\\
&=&{(m-m_0)^2\over 4G}-{N_c m^3\over4\pi^2}\left[\Lambda\Big(1+{2\Lambda^2\over m^2}\Big)\sqrt{1+{\Lambda^2\over m^2}}-m\ln\left({\Lambda\over m}
+\sqrt{1+{\Lambda^2\over m^2}}\right)\right]+{2N_c\over\pi^2}T\sum_{s=\pm}\int_0^\infty p^2 dp \ln\Big(1+e^{(E(p)+s\mu)/T}\Big)
\nonumber\\&&+{N_{c}\over8\pi^2}\sum_{\rm f=u,d}\int_0^\infty {ds\over s^3}e^{-m^2s}\Big[\vartheta_3\left({\pi\over 2}+i{\mu\over 2T},e^{-\Big|{q_{\rm f}E\over 4\tan(q_{\rm f}Es)T^2}\Big|}\right){{q_{\rm f}Es\over\tan(q_{\rm f}Es)}}
-\vartheta_3\left({\pi\over 2}+i{\mu\over 2T},e^{-{1\over 4sT^2}}\right)\Big].
\end{eqnarray}
\end{widetext}
The last term is still logarithmically divergent for the integral domain around $s=0$. This is not a real problem because what we really care is the difference $\Omega(m)-\Omega(0)$ for different phases which is of course convergent. It should be noticed that the pure magnetic correspondence of $\Omega(m)-\Omega(0)$ has the same magnetic field dependent part as that in Ref.~\cite{Allen:2015paa} in the zero temperature and chemical potential limit, which justifies our present regularization scheme in the presence of electromagnetic field.

For pure electric field case, we find that there are infinite poles in the thermodynamic potential and the gap equation due to the presence of the electric-field-related tangent term which are from the contribution of Schwinger pair production. These poles render the thermodynamic potential to be complex with its real part determining the ground state while the imaginary part giving the Schwinger pair production rate. Hereafter we will simply call the real part the thermodynamic potential. We give explicit expression for the Schwinger production rate, that is, the probability per unit time and per unit volume for pair production~\cite{Schwinger:1951nm},
\begin{eqnarray}
\Gamma=-2{\rm Im}\;\Omega(m)=\sum_{\rm f=u,d}\sum_{n=1}^\infty{N_c(q_{\rm f}E)^2\over4\pi}{e^{-n\pi m^2/|q_{\rm f}E|}\over(n\pi)^2},
\end{eqnarray}
which does not depend on the temperature $T$ or chemical potential $\mu$ explicitly. However, since the quark mass $m$ depends on both $T$ and $\mu$ as can be seen in the gap equation, they will affect the pair production rate implicitly through $m$. The numerical simulations will be presented in next section.

As well known, in the chiral symmetry breaking phase, $\sigma$ and $\pi_0$ modes are massive and massless collective excitations, respectively. The mass of $\sigma$ mode is very hard to determine with Schwinger's approach because of the difficulty in summing out the Matsubara frequency analytically as we have mentioned and the awful feature of the function $|\tan(q_{\rm f}Es)|$ in the exponential. Therefore, we merely care about the expansion coefficients of the inverse propagator of $\pi_0$ mode around small ${\bf q}$ and their relative ratio with each other. Then, in the case of pure electric field, by taking the limit $B\rightarrow0$, the inverse propagator of $\pi_0$ can be easily derived from Eq. (\ref{propagator}) as the following:
\begin{widetext}
\begin{eqnarray}
i\tilde{\cal D}_0^{-1}(q)
&=&\frac{1}{2G}-4N_{c}\sum_{\rm f=u,d}\sum_{\tilde{p}}\int_0^\infty {ds}\int_0^\infty {ds'}\exp\Big\{-m^2(s+s')-{{s}[(p_1+q_1)^2+(p_2+q_2)^2]}-{{s'}(p_1^2+p_2^2)}\nonumber\\
&&-{\tan(q_{\rm f}Es)[(\tilde{p}_4+q_4)^2+(p_3+q_3)^2]\over q_{\rm f}E}-{\tan(q_{\rm f}Es')(\tilde{p}_4^2+p_3^2)\over q_{\rm f}E}\Big\}\Big\{\big[m^2+({p}_2+q_2){p}_2+({p}_1+q_1){p}_1\big]\nonumber\\
&&\times\big(1-\tan(q_{\rm f}Es)\tan(q_{\rm f}Es')\big)+\big[(\tilde{p}_4+q_4)\tilde{p}_4+({p}_3+q_3){p}_3\big]
\big(1+\tan^2(q_{\rm f}Es)\big)\big(1+\tan^2(q_{\rm f}Es')\big)\Big\}.
\end{eqnarray}
First, we regularize the divergence of $i\tilde{\cal D}_0^{-1}(q)$ with the three-momentum cutoff as mentioned in Ref.~\cite{Cao:2014uva,Cao:2015xja} and then by taking into account the gap equation, the inverse propagator becomes
\begin{eqnarray}
i\tilde{\cal D}_0^{-1}(q)
&=&-4N_{c}\sum_{\rm f=u,d}\sum_{\tilde{p}}\int_0^\infty {ds}\int_0^\infty {ds'}\exp\Big\{-m^2(s+s')-{{s}[(p_1+q_1)^2+(p_2+q_2)^2]}-{{s'}(p_1^2+p_2^2)}\nonumber\\
&&-{\tan(q_{\rm f}Es)[(\tilde{p}_4+q_4)^2+(p_3+q_3)^2]\over q_{\rm f}E}-{\tan(q_{\rm f}Es')(\tilde{p}_4^2+p_3^2)\over q_{\rm f}E}\Big\}\Big\{\big[m^2+({p}_2+q_2){p}_2+({p}_1+q_1){p}_1\big]\nonumber\\
&&\big(1-\tan(q_{\rm f}Es)\tan(q_{\rm f}Es')\big)+\big[(\tilde{p}_4+q_4)\tilde{p}_4+({p}_3+q_3){p}_3\big]
\big(1+\tan^2(q_{\rm f}Es)\big)\big(1+\tan^2(q_{\rm f}Es')\big)\nonumber\\
&&-\Big[m^2+({p}_2+q_2)^2+({p}_1+q_1)^2\Big]-\Big[({p}_4+q_4)^2+({p}_3+q_3)^2\Big](1+\tan^2(q_{\rm f}Es))\Big\},
\end{eqnarray}
which we have verified to be vanishing at $q=0$ numerically thus the Goldstone theorem is satisfied. In $3+1$ dimensions, this regularized inverse propagator is still divergent at finite $q$. To keep consistent with the Goldstone theorem, we shift the divergence to an electric-field-independent part which is then evaluated with the three-momentum cutoff, that is,
\begin{eqnarray}
i\tilde{\cal D}_0^{-1}(q)&=&\Big[\Pi_0^E(q)-\Pi_0^0(q)\Big]+\Pi_0^0(q,\Lambda).
\end{eqnarray}
The expansion of $\Pi_0^0(q,\Lambda)$ around small $q$ was evaluated before~\cite{Klevansky:1992qe} and has a simple form:
\begin{eqnarray}
&&\ \ \ \ \ \ \ \ \ \ \ \ \Pi_0^0(q,\Lambda)=\zeta_4q_4^2+\zeta_{i}{\mathbf q}^2+o(q^3),\nonumber\\
\zeta_4&=&\!N_c\!\!\!\int^\Lambda\!\!\!{d^3p\over(2\pi)^3}{1\over E^3({p})}\!\!-N_c\!\sum_{s=\pm}\!\int\!\!\!{d^3p\over(2\pi)^3}{1\over E^3({p})(1+e^{(E({p})+s\mu)/ T})},\nonumber\\
\zeta_{i}&=&\!N_c\!\!\!\int^\Lambda\!\!\!{d^3p\over(2\pi)^3}{1\over E^3({p})}-\!N_c\!\sum_{s=\pm}\!\!\int\!\!\!{d^3p\over(2\pi)^3}\Big[{1\over E^3({p})(1+e^{(E({p})+s\mu)/ T})}\!+\!{{\rm sech}^2\big({E({p})+s\mu\over2T}\big)\over 4TE^2({p})}\Big].
\end{eqnarray}
And the expansion coefficients $\xi^E$ of $\Pi_0^E(q)$ around small ${\bf q}$ can be derived directly from (\ref{xi}) as
\begin{eqnarray}
\xi_{\bot}^E&=&{N_{c}\over4\pi^{2}}\sum_{\rm f=u,d}\int_0^\infty {ds}\int_0^\infty {ds'}
e^{-(s+s')m^2}
{1\over s+s'}{q_{\rm f}E\over \tan(q_{\rm f}Es)+\tan(q_{\rm f}Es')}\vartheta_3\left({\pi\over 2}+i{\mu\over 2T},\exp\Big(-\Big|{q_{\rm f}E\over 4(\tan(q_{\rm f}Es)+\tan(q_{\rm f}Es'))T^2}\Big|\Big)\right)
\nonumber\\
&&\left\{\!{ss'\over (s+s')}\Big[\!\!-\!m^2\tan(q_{\rm f}Es')\Big(\tan(q_{\rm f}Es)\!+\!\tan(q_{\rm f}Es')\Big)\!+\!{{s'}-s\over s(s+s')}\tan^2(q_{\rm f}Es')
\Big]\!-\!{2ss'\over (s+s')^2}\tan(q_{\rm f}Es)\tan(q_{\rm f}Es')\!+\!{{s'}\over s+s'}\right\},\\
\xi_{3}^E&=&{N_{c}\over4\pi^{2}}\sum_{\rm f=u,d}\int_0^\infty {ds}\int_0^\infty {ds'}
e^{-(s+s')m^2}
{1\over s+s'}{q_{\rm f}E\over \tan(q_{\rm f}Es)+\tan(q_{\rm f}Es')}\vartheta_3\left({\pi\over 2}+i{\mu\over 2T},\exp\Big(-\Big|{q_{\rm f}E\over 4(\tan(q_{\rm f}Es)+\tan(q_{\rm f}Es'))T^2}\Big|\Big)\right)
\nonumber\\
&&\left\{-{\tan(q_{\rm f}Es)\tan^2(q_{\rm f}Es')\over q_{\rm f}E}(m^2+{1\over s+s'})
+{\tan(q_{\rm f}Es')\over \tan(q_{\rm f}Es)+\tan(q_{\rm f}Es')}\Big(1+\tan^2(q_{\rm f}Es)\Big)\Big(1+\tan^2(q_{\rm f}Es')\Big)
\right\}.
\end{eqnarray}
Finally, by integrating these results together, we have a regularized form for the small $q$ expansion of the inverse propagator
\begin{eqnarray}
i\tilde{\cal D}_0^{-1}(q)&=&(\xi_4^E-\xi_4^0+\zeta_4)q_4^2+(\xi_3^E-\xi_3^0+\zeta_{i})q_3^2+(\xi_\bot^E-\xi_\bot^0+\zeta_{i})
(q_1^2+q_2^2)+o(q^3),
\end{eqnarray}
\end{widetext}
where $\xi_4^E-\xi_4^0$ is only formal but equals to $\xi_3^E-\xi_3^0$ at zero temperature. Since $\xi_4^E-\xi_4^0$ themselves are hard to handle within Schwinger's approach at finite temperature, we will turn to study the relative ratio between sound velocities along different directions instead, that is,
\begin{eqnarray}
v_{\bot}/v_3=\sqrt{\xi_\bot^E-\xi_\bot^0+\zeta_{i}\over\xi_3^E-\xi_3^0+\zeta_{i}}.
\end{eqnarray}
As the longitudinal velocity $v_3=1$ at zero temperature and vanishing chemical potential, this ratio becomes the true sound velocity in the transverse directions in that case.

\section{Numerical calculations}\label{sec2}

\subsection{The case at $T=0$}
With the establishment of operable formalism, we now present our numerical calculations in the following.
We start with the simplest case at vanishing temperature. As has been mentioned before, the evolution of the dynamical mass $m$ with electric field $E$ had already been studied in this case but only one flavor was concerned~\cite{Klevansky:1989vi}. The calculations here also include the velocity ratio of Goldstone modes and the Schwinger pair production rate. At zero temperature, only the case with vanishing chemical potential can be studied within Schwinger's approach as will be shown in Sec.\ref{Tn0}. Then the gap equation is further reduced to
\begin{eqnarray}
{m-m_0\over 2G}&=&{N_c m^2\over\pi^2}\left[\Lambda\sqrt{1+{\Lambda^2\over m^2}}-m\ln\left({\Lambda\over m}
+\sqrt{1+{\Lambda^2\over m^2}}\right)\right]\nonumber\\
&&+{mN_{c}\over4\pi^2}\sum_{\rm f=u,d}\int_0^\infty {ds\over s^2}e^{-m^2s}\Big[{{q_{\rm f}Es\over\tan(q_{\rm f}Es)}}
-1\Big].
\end{eqnarray}
And after several steps of manipulation, the expansion coefficients $\xi^E$ are found to be very simple:
\begin{eqnarray}
\xi_\bot^E&=&{N_{c}\over8\pi^{2}}\sum_{\rm f=u,d}\int_0^\infty {ds}
e^{-sm^2}{q_{\rm f}E\over \tan(q_{\rm f}Es)},\\
\xi_3^E&=&{N_{c}\over4\pi^{2}}\int_0^\infty {ds}
{e^{-sm^2}\over s}.
\end{eqnarray}
It is interesting to see that $\xi_\bot$ depends on the electric field explicitly but $\xi_3$ does not, compared to the opposite feature in the pure magnetic field case~\cite{Gusynin:1994re,Gusynin:1994xp,Gusynin:1995nb}. This is another manifestation of the duality between the electric and magnetic fields. Note that the integrations are different from those in Ref.~\cite{Gusynin:1995nb} by $1/3$ when introducing infrared cutoff $1/\Lambda^2$ for the integration of $s$ because we take partial integrals before regularization. However, this makes no difference in our regularization scheme as $1/3$ cancels in $\xi_\bot^E-\xi_\bot^0$ and $\xi_3^E-\xi_3^0$.

\begin{figure}[!htb]
\begin{center}
\includegraphics[width=8cm]{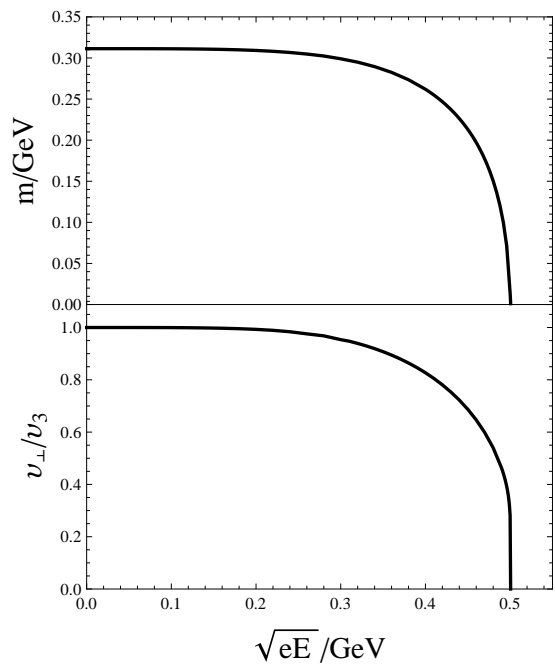}
\caption{The dependence of the quark mass $m$ and the sound velocity ratio $v_\bot/v_3$ on the electric field $\sqrt{eE}$ at $T=0$.\label{fig1}}
\end{center}
\end{figure}

We choose the parameter set $G=5.01{\rm GeV}^{-2}$ and $\Lambda=0.65 {\rm GeV}$ in chiral limit~\cite{Zhuang:1994dw} and the numerical results are shown in Fig.\ref{fig1}. As can be seen, the decreasing feature of fermion mass with the electric field as shown in Ref.~\cite{Klevansky:1989vi} is well reproduced which indicates the inverse catalysis effect of the pure electric field. When the electric field is larger than a critical value $E_c$ the chiral symmetry is restored via a seemly second order phase transition. The sound velocity in the transverse directions is also found to decrease with the electric field and vanish when $E$ becomes very close to the critical electric field $E_c$ for the chiral symmetry restoration. Actually, the decreasing at small $E$ is inevitable because of the constraining of causality when the start point is at the velocity of light, and the underlying physics is that the longitudinal motion is more favored for charged quarks and antiquarks when an electric field exists. So we get a general result that the longitudinal motion is always much more free for Goldstone mode in the presence of either an electric field or a magnetic field. The vanishing of the sound velocity near $E_c$ can be shown analytically. At the critical electric field, the transverse expansion coefficient becomes
\begin{eqnarray}
\xi_\bot^E-\xi_\bot^0&=&{N_{c}\over8\pi^{2}}\sum_{\rm f=u,d}\int_0^\infty {ds}
\Big({q_{\rm f}E\over \tan(q_{\rm f}Es)}-{1\over s}\Big),\\
&{\propto}&\ln\Big|{\sin(q_{\rm f}Es)\over s}\Big|_{s\rightarrow\infty}\rightarrow-\infty.
\end{eqnarray}
Thus, at some point below the critical electric field, we will find $\xi_\bot^E-\xi_\bot^0+\zeta_{i}=0$ and the velocity $v_\bot$ vanishes. The numerical calculation shows that this point is very near the critical electric field.

\begin{figure}[!htb]
\begin{center}
\includegraphics[width=8cm]{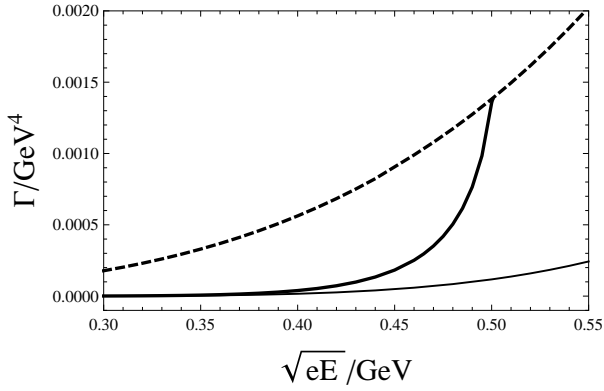}
\caption{The dependence of the Schwinger pair production rate $\Gamma$ on the electric field $\sqrt{eE}$ at $T=0$. The thick solid line is the real case, and the thin and the dashed lines correspond to the limiting cases of $m=0.311{\rm GeV}$ and $m=0$, respectively.\label{fig2}}
\end{center}
\end{figure}

The pair production rate is also evaluated as shown in Fig.\ref{fig2}.
In order to show explicitly how chiral symmetry restoration changes the feature of pair production, we also include the limit where $m$
is kept to be a constant $m=0.311{\rm GeV}$ as that at $E=0$ and the limit where $m$ is always zero. As can be seen, the production rate is enhanced by one order at the critical electric field comparing to the $m=0.311{\rm GeV}$ case. If $m=0$, the production rate is simply a quadratic function of the electric field, that is,
\begin{eqnarray}
\Gamma=\sum_{\rm f=u,d}{N_c(q_{\rm f}E)^2\over24\pi}.
\end{eqnarray}
Of course, the real $\Gamma$ with the $E$-dependent quark mass coincides with this power law feature when $E>E_c$.

It is illuminative to estimate the generated number of charged quark-antiquark pairs in the quark-gluon plasma (QGP) produced in HICs. Because in the QGP phase the chiral symmetry is restored, the dynamical quark mass can be assumed to be zero. According to the recent numerical simulations, the electric fields generated in Au + Au collisions at $\sqrt{s}=200$ GeV owing to fluctuation is of the order $eE\sim m_\pi^2$ while in Pb + Pb collisions at $\sqrt{s}=2.76$ TeV is of the order $eE\sim 20 m_\pi^2$~\cite{Deng:2012pc}. Then supposing the space-time volume of the QGP is of the order $(5{\rm fm})^4$, the total pair production number is
\begin{eqnarray}
N_{\rm RHIC}&\sim&{5\over 72\pi}*\Big({5*0.14\over 0.197}\Big)^4\approx3.5,\\
N_{\rm LHC}&\sim&{5\over 72\pi}*\Big({22*0.14\over 0.197}\Big)^4\approx1400,
\end{eqnarray}
This indicates that the Schwinger pair production of quark and antiquark is actually quite significant in HICs.

\subsection{The case with $T\neq0$ and $\mu\neq0$}\label{Tn0}
\begin{figure}[!htb]
\begin{center}
\includegraphics[width=8cm]{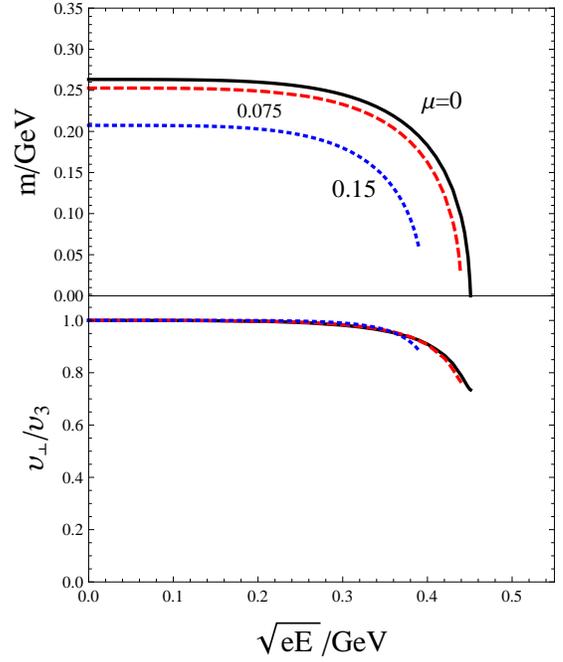}
\caption{(Color online) The quark mass $m$ and the sound velocity ratio $v_\bot/v_3$ as functions of the electric field $\sqrt{eE}$ at $T=0.13{\rm GeV}$ for different baryon chemical potentials $\mu=0$ (black solid line), $0.075$ (red dashed line) and $0.15{\rm GeV}$ (blue dotted line).\label{fig3}}
\end{center}
\end{figure}

In order to show the effect of temperature, we choose to work at $T=0.13{\rm GeV}$ which is below the critical temperature $T_c=0.19{\rm GeV}$ at $E=0$ in NJL model~\cite{Zhuang:1994dw}. The numerical results for quark mass $m$ and the sound velocity ratio $v_\bot/v_3$ are shown in Fig.\ref{fig3} for baryon chemical potential $\mu=0,\ 0.075$ and $0.15{\rm GeV}$ and temperature $T=0.13{\rm GeV}$. As had been stated in our previous section, the Schwinger's approach is valid only when the magnitude of $\mu$ is constrained a region so that the integral of the proper time $s$ is convergent. The fermion Matsubara frequency is $\omega_n=(2n+1)\pi T$ and ${\rm Re}(\tilde{p}_4^2)$ is guaranteed to be positive only if $\pi T>\mu$. Here, the effective range of $\mu$ is solely determined by the temperature which is different from the pure magnetic field case where only $m^2+(\pi T)^2>\mu^2$ has to be satisfied. The main reason is that in the pure electric field case, there is a tangent function in front of $\tilde{p}_4^2$ which can become infinite at finite $s$ and make the mass term irrelevant. That is why we choose the temperature to be large so that we can determine the critical chemical potential when small electric field is present.

Comparing Fig.\ref{fig3} together with Fig.\ref{fig1}, we can find that the temperature $T$ and baryon chemical potential $\mu$ both tend to melt the chiral condensate as we expected. The curves for finite chemical potential cases stop at certain points where first-order phase transitions occur. And the sound velocity ratio is almost $1$ up to $\sqrt{eE}=0.3{\rm GeV}$ and then it is found to decrease faster for larger chemical potentials. This can be easily explained by looking at the behavior of $m$: At larger chemical potentials, $m$ is smaller and decreases faster with $\sqrt{eE}$; thus, the effect of electric field shows up earlier in the velocity ratio. Because the finite temperature would make the integral of proper time finite over the ultraviolet region, the velocity ratio doesn't decrease to zero even though $m=0$ at the critical electric field even for the vanishing chemical potential case.

\begin{figure}[!htb]
\begin{center}
\includegraphics[width=8cm]{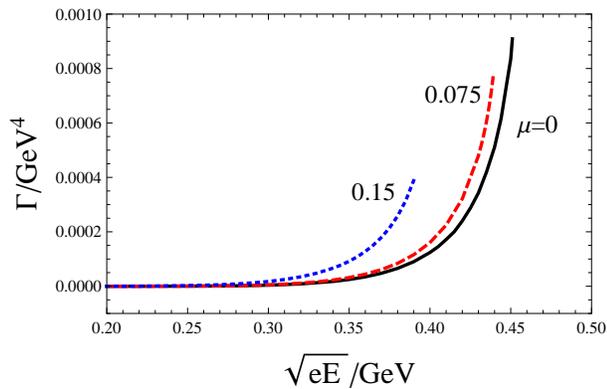}
\caption{(Color online) The pair production rate $\Gamma$ as a function of the electric field $\sqrt{eE}$ at $T=0.13{\rm GeV}$ for different baryon chemical potentials $\mu=0$ (black solid line), $0.075$ (red dashed line) and $0.15{\rm GeV}$ (blue dotted line).\label{fig4}}
\end{center}
\end{figure}

The pair production rate is also calculated as illuminated in Fig.\ref{fig4}.
The plot shows that the production rate becomes larger for larger chemical potentials which is consistent with the fact that $m$ decreases with the baryon chemical potential.

\begin{figure}[!htb]
\begin{center}
\includegraphics[width=8cm]{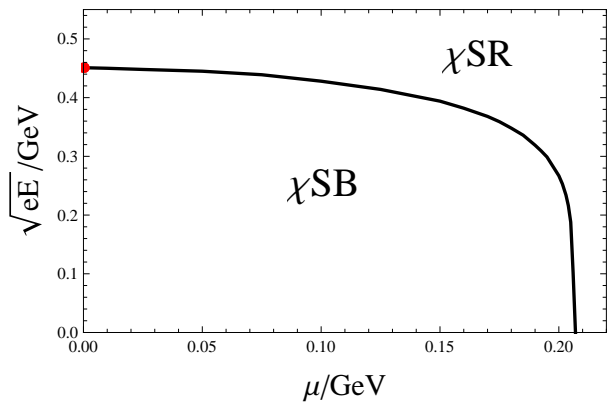}
\caption{The $E-\mu$ phase diagram at $T=0.13{\rm GeV}$. The regions inside and outside the transition line correspond to chiral symmetry breaking ($\chi$SB) and chiral symmetry restoration ($\chi$SR) phases, respectively. The red point on the vertical axis stands for a second-order transition.\label{fig5}}
\end{center}
\end{figure}

Finally, we present the $E-\mu$ phase diagram in Fig.\ref{fig5}. The critical electric field decreases with chemical potential as expected. As the dynamical mass $m$ drops fast around $\mu=0.2{\rm GeV}$ at $E=0$, the critical electric field also drops sharply there. The phase transition is found be of second order at vanishing chemical potential and first order at finite chemical potential which implies that the red point on the vertical axis is actually a critical point.

\section{Summary}\label{sec3}
In summary, we have established a general formalism for chiral symmetry breaking and restoration in the presence of parallel electric and magnetic fields. A modified regularization scheme is introduced to treat the divergence in the gap equation and the expansion coefficients of Goldstone modes consistently in the pure electric field case which is also valid in the general case with both electric and magnetic fields. We then focused on the pure electric field case.

Firstly, the effects of a pure electric field to dynamical mass, velocity ratio and pair production rate were systematically explored. The previous qualitative result that chiral symmetry will be restored by electric field was well reproduced at zero temperature. And the decreasing of the sound velocity ratio with electric field indicates that the longitudinal motion is much favored than the transverse motion at large $E$. At a given finite temperature, the anisotropy of motion shows up much earlier for larger chemical potential due to chiral symmetry restoration. The pair production rate is also evaluated consistently with chiral symmetry restoration and found to be greatly enhanced near the transition point. The number of pair production in the QGP systems created in HICs is roughly estimated to be considerably large whose dynamical consequence is worth of further investigating. Finally, the $E-\mu$ phase diagram is given at fixed finite temperature. The chiral symmetry breaking and restoration transition is found to be of first order for any finite value of $\mu$ and the $\mu=0$ point can be identified as a critical point.

The more general case with non-vanishing parallel electric and magnetic fields but zero temperature and chemical potentials was also studied in the NJL model recently which is much more involved~\cite{Cao:2015cka}. The case with finite temperature and chemical potential is in progress. Furthermore, at finite temperature, there is no Lorentz invariance for the systems and the effects of the electromagnetic configuration with orthogonal electric and magnetic fields can not be described simply by the Lorentz invariants $I_1=B^2-E^2$. We have preliminarily studied this case recently and the formalism seems quite complicated. A lot of work still need to be done to explore the effects of a general configuration of electromagnetic field to the chiral symmetry breaking and restoration and the related physical phenomena, especially when the temperature and chemical potential are nonzero.

In the present paper, we have focused on the chiral phase transtion. We however note that it will be also interesting to explore the possible electric field effects on the deconfinment phase transition and the modification of the asymptotic freedom phenomenon of QCD, as it is natural to expect that a large $E$ will break the quark-antiquark pair bounded in mesons which may lead to deconfinement and also provide an energy scale along which the coupling constant of QCD may run. We leave these studies as future tasks.

\acknowledgments
GC and XGH are supported by Shanghai Natural Science Foundation (Grant No. 14ZR1403000), the Key Laboratory of Quark and Lepton Physics (MOE) of CCNU (Grant No. QLPL20122), the Young 1000 Talents Program of China, and Scientific Research Foundation of State Education Ministry for Returned Scholars.

\begin{widetext}
\appendix
\section{The neutral collective modes}\label{nmodes}
In this appendix, we'd like to derive the general forms of the inverse propagators of neutral collective modes explicitly by following (\ref{boson1}). For $\sigma$ mode, the inverse propagator becomes
\begin{eqnarray}
i\tilde{\cal D}_\sigma^{-1}(q)&=&\frac{1}{2G}-N_{c}\sum_{\rm f=u,d}\sum_{\tilde{p}}\int_0^\infty {ds}\int_0^\infty {ds'}\exp\Big\{-im^2s-{i[(\tilde{p}_4+q_4)^2+(p_3+q_3)^2]\over q_{\rm f}E\coth(q_{\rm f}Es)}-{i[(p_1+q_1)^2+(p_2+q_2)^2]\over q_{\rm f}B\cot(q_{\rm f}Bs)}\nonumber\\
&&-im^2s'-{i(\tilde{p}_4^2+p_3^2)\over q_{\rm f}E\coth(q_{\rm f}Es')}-{i(p_1^2+p_2^2)\over q_{\rm f}B\cot(q_{\rm f}Bs')}\Big\}\;tr\Big[m-\gamma^4(\tilde{p}_4+q_4-{i\tanh(q_{\rm f}Es)}(p_3+q_3))-\gamma^3({p}_3+q_3\nonumber\\
&&+{i\tanh(q_{\rm f}Es)}(\tilde{p}_4+q_4))-\gamma^1(p_1+q_1-{\tan(q_{\rm f}Bs)}(p_2+q_2))-\gamma^2(p_2+q_2+{\tan(q_{\rm f}Bs)}(p_1+q_1))\Big]\nonumber\\
&&\Big[1+i\tan(q_{\rm f}Bs)\tanh(q_{\rm f}Es)\gamma^5+\tan(q_{\rm f}Bs)\gamma^1\gamma^2+i\tanh(q_{\rm f}Es)\gamma^4\gamma^3\Big]\Big[m-\gamma^4(\tilde{p}_4-{i\tanh(q_{\rm f}Es')}p_3)\nonumber\\
&&-\gamma^3({p}_3+{i\tanh(q_{\rm f}Es')}\tilde{p}_4)-\gamma^1(p_1-{\tan(q_{\rm f}Bs')}p_2)-\gamma^2(p_2+{\tan(q_{\rm f}Bs')}p_1)\Big]\Big[1+i\tan(q_{\rm f}Bs')\tanh(q_{\rm f}Es')\gamma^5\nonumber\\
&&+\tan(q_{\rm f}Bs')\gamma^1\gamma^2+i\tanh(q_{\rm f}Es')\gamma^4\gamma^3\Big]
\end{eqnarray}
by substituting the effective propagators of quarks. Then the trace terms can be evaluated as the following:
\begin{eqnarray}
tr_1&=&tr\Big[m-\gamma^4(\tilde{p}_4+q_4-{i\tanh(q_{\rm f}Es)}(p_3+q_3))-\gamma^3({p}_3+q_3+{i\tanh(q_{\rm f}Es)}(\tilde{p}_4+q_4))-\gamma^1(p_1+q_1-{\tan(q_{\rm f}Bs)}(p_2+q_2))\nonumber\\
&&\;\;\;-\gamma^2(p_2+q_2+{\tan(q_{\rm f}Bs)}(p_1+q_1))\Big]\Big[1+i\tan(q_{\rm f}Bs)\tanh(q_{\rm f}Es)\gamma^5+\tan(q_{\rm f}Bs)\gamma^1\gamma^2+i\tanh(q_{\rm f}Es)\gamma^4\gamma^3\Big]\nonumber\\
&&\;\;\;\Big[m-\gamma^4(\tilde{p}_4-{i\tanh(q_{\rm f}Es')}p_3)-\gamma^3({p}_3+{i\tanh(q_{\rm f}Es')}\tilde{p}_4)-\gamma^1(p_1-{\tan(q_{\rm f}Bs')}p_2)-\gamma^2(p_2+{\tan(q_{\rm f}Bs')}p_1)\Big]\nonumber\\
&=&4\Big\{m^2\!-\!(\tilde{p}_4\!+\!q_4\!-\!{i\tanh(q_{\rm f}Es)}(p_3\!+\!q_3))(\tilde{p}_4\!-\!{i\tanh(q_{\rm f}Es')}p_3)\!-\!({p}_3\!+\!q_3\!+\!{i\tanh(q_{\rm f}Es)}(\tilde{p}_4\!+\!q_4))({p}_3\!+\!{i\tanh(q_{\rm f}Es')}\tilde{p}_4)\nonumber\\
&&\;\;\;-(p_1+q_1-{\tan(q_{\rm f}Bs)}(p_2+q_2))(p_1-{\tan(q_{\rm f}Bs')}p_2)-(p_2+q_2+{\tan(q_{\rm f}Bs)}(p_1+q_1))(p_2+{\tan(q_{\rm f}Bs')}p_1)\nonumber\\
&&\;\;\;+{\tan(q_{\rm f}Bs)}\big[(p_1+q_1-{\tan(q_{\rm f}Bs)}(p_2+q_2))(p_2+{\tan(q_{\rm f}Bs')}p_1)-(p_2+q_2+{\tan(q_{\rm f}Bs)}(p_1+q_1))(p_1-{\tan(q_{\rm f}Bs')}p_2)\big]\nonumber\\
&&\;\;\;+i\tanh(q_{\rm f}Es)\big[(\tilde{p}_4+q_4-{i\tanh(q_{\rm f}Es)}(p_3+q_3))({p}_3+{i\tanh(q_{\rm f}Es')}\tilde{p}_4)-({p}_3+q_3+{i\tanh(q_{\rm f}Es)}(\tilde{p}_4+q_4))\nonumber\\
&&\;\;\;(\tilde{p}_4-{i\tanh(q_{\rm f}Es')}p_3)\big]\Big\},\\
tr_2&=&tr\Big[m-\gamma^4(\tilde{p}_4+q_4-{i\tanh(q_{\rm f}Es)}(p_3+q_3))-\gamma^3({p}_3+q_3+{i\tanh(q_{\rm f}Es)}(\tilde{p}_4+q_4))-\gamma^1(p_1+q_1-{\tan(q_{\rm f}Bs)}(p_2+q_2))\nonumber\\
&&\;\;\;-\gamma^2(p_2+q_2+{\tan(q_{\rm f}Bs)}(p_1+q_1))\Big]\Big[1+i\tan(q_{\rm f}Bs)\tanh(q_{\rm f}Es)\gamma^5+\tan(q_{\rm f}Bs)\gamma^1\gamma^2+i\tanh(q_{\rm f}Es)\gamma^4\gamma^3\Big]\nonumber\\
&&\;\;\;\Big[m-\gamma^4(\tilde{p}_4-{i\tanh(q_{\rm f}Es')}p_3)-\gamma^3({p}_3+{i\tanh(q_{\rm f}Es')}\tilde{p}_4)-\gamma^1(p_1-{\tan(q_{\rm f}Bs')}p_2)-\gamma^2(p_2+{\tan(q_{\rm f}Bs')}p_1)\Big]\nonumber\\
&&\;\;\;i\tan(q_{\rm f}Bs')\tanh(q_{\rm f}Es')\gamma^5\nonumber\\
&=&4\tan(q_{\rm f}Bs')\tanh(q_{\rm f}Es')\Big\{-\tan(q_{\rm f}Bs)\tanh(q_{\rm f}Es)\big[m^2+(\tilde{p}_4+q_4-{i\tanh(q_{\rm f}Es)}(p_3+q_3))(\tilde{p}_4-{i\tanh(q_{\rm f}Es')}p_3)\nonumber\\
&&\;\;\;+({p}_3+q_3+{i\tanh(q_{\rm f}Es)}(\tilde{p}_4+q_4))({p}_3+{i\tanh(q_{\rm f}Es')}\tilde{p}_4)+(p_1+q_1-{\tan(q_{\rm f}Bs)}(p_2+q_2))(p_1-{\tan(q_{\rm f}Bs')}p_2)\nonumber\\
&&\;\;\;+(p_2\!+\!q_2\!+\!{\tan(q_{\rm f}Bs)}(p_1\!+\!q_1))(p_2\!+\!{\tan(q_{\rm f}Bs')}p_1)\big]\!+\!i{\tan(q_{\rm f}Bs)}\big[(\tilde{p}_4\!+\!q_4\!-\!{i\tanh(q_{\rm f}Es)}(p_3\!+\!q_3))({p}_3\!+\!{i\tanh(q_{\rm f}Es')}\tilde{p}_4)\nonumber\\
&&\;\;\;-({p}_3\!+\!q_3\!+\!{i\tanh(q_{\rm f}Es)}(\tilde{p}_4\!+\!q_4))(\tilde{p}_4\!-\!{i\tanh(q_{\rm f}Es')}p_3)\big]\!-\!\tanh(q_{\rm f}Es)\big[(p_1\!+\!q_1\!-\!{\tan(q_{\rm f}Bs)}(p_2\!+\!q_2))(p_2\!+\!{\tan(q_{\rm f}Bs')}p_1)\nonumber\\
&&\;\;\;-(p_2+q_2+{\tan(q_{\rm f}Bs)}(p_1+q_1))(p_1-{\tan(q_{\rm f}Bs')}p_2)\big]\Big\},\\
tr_3&=&tr\Big[m-\gamma^4(\tilde{p}_4+q_4-{i\tanh(q_{\rm f}Es)}(p_3+q_3))-\gamma^3({p}_3+q_3+{i\tanh(q_{\rm f}Es)}(\tilde{p}_4+q_4))-\gamma^1(p_1+q_1-{\tan(q_{\rm f}Bs)}(p_2+q_2))\nonumber\\
&&\;\;\;-\gamma^2(p_2+q_2+{\tan(q_{\rm f}Bs)}(p_1+q_1))\Big]\Big[1+i\tan(q_{\rm f}Bs)\tanh(q_{\rm f}Es)\gamma^5+\tan(q_{\rm f}Bs)\gamma^1\gamma^2+i\tanh(q_{\rm f}Es)\gamma^4\gamma^3\Big]\nonumber\\
&&\;\;\;\Big[m\!-\!\gamma^4(\tilde{p}_4\!-\!{i\tanh(q_{\rm f}Es')}p_3)\!-\!\gamma^3({p}_3\!+\!{i\tanh(q_{\rm f}Es')}\tilde{p}_4)\!-\!\gamma^1(p_1\!-\!{\tan(q_{\rm f}Bs')}p_2)\!-\!\gamma^2(p_2\!+\!{\tan(q_{\rm f}Bs')}p_1)\Big]\tan(q_{\rm f}Bs')\gamma^1\gamma^2\nonumber\\
&=&4\tan(q_{\rm f}Bs')\Big\{-\tan(q_{\rm f}Bs)\big[m^2-(\tilde{p}_4+q_4-{i\tanh(q_{\rm f}Es)}(p_3+q_3))(\tilde{p}_4-{i\tanh(q_{\rm f}Es')}p_3)-({p}_3+q_3+{i\tanh(q_{\rm f}Es)}\nonumber\\
&&\;\;\;-(\tilde{p}_4+q_4))({p}_3+{i\tanh(q_{\rm f}Es')}\tilde{p}_4)+(p_1+q_1-{\tan(q_{\rm f}Bs)}(p_2+q_2))(p_1-{\tan(q_{\rm f}Bs')}p_2)+(p_2+q_2+{\tan(q_{\rm f}Bs)})\nonumber\\
&&\;\;\;(p_1+q_1))(p_2+{\tan(q_{\rm f}Bs')}p_1)\big]-i{\tan(q_{\rm f}Bs)}{\tanh(q_{\rm f}Es)}\big[(\tilde{p}_4+q_4-{i\tanh(q_{\rm f}Es)}(p_3+q_3))({p}_3+{i\tanh(q_{\rm f}Es')}\tilde{p}_4)\nonumber\\
&&\;\;\;-({p}_3+q_3+{i\tanh(q_{\rm f}Es)}(\tilde{p}_4+q_4))(\tilde{p}_4-{i\tanh(q_{\rm f}Es')}p_3)\big]-\big[(p_1+q_1-{\tan(q_{\rm f}Bs)}(p_2+q_2))(p_2+{\tan(q_{\rm f}Bs')}p_1)\nonumber\\
&&\;\;\;-(p_2+q_2+{\tan(q_{\rm f}Bs)}(p_1+q_1))(p_1-{\tan(q_{\rm f}Bs')}p_2)\big]\Big\},
\end{eqnarray}
\begin{eqnarray}
tr_4&=&tr\Big[m-\gamma^4(\tilde{p}_4+q_4-{i\tanh(q_{\rm f}Es)}(p_3+q_3))-\gamma^3({p}_3+q_3+{i\tanh(q_{\rm f}Es)}(\tilde{p}_4+q_4))-\gamma^1(p_1+q_1-{\tan(q_{\rm f}Bs)}(p_2+q_2))\nonumber\\
&&\;\;\;-\gamma^2(p_2+q_2+{\tan(q_{\rm f}Bs)}(p_1+q_1))\Big]\Big[1+i\tan(q_{\rm f}Bs)\tanh(q_{\rm f}Es)\gamma^5+\tan(q_{\rm f}Bs)\gamma^1\gamma^2+i\tanh(q_{\rm f}Es)\gamma^4\gamma^3\Big]\nonumber\\
&&\;\;\;\Big[m\!-\!\gamma^4(\tilde{p}_4\!-\!{i\tanh(q_{\rm f}Es')}p_3)\!-\!\gamma^3({p}_3\!+\!{i\tanh(q_{\rm f}Es')}\tilde{p}_4)\!-\!\gamma^1(p_1\!-\!{\tan(q_{\rm f}Bs')}p_2)\!-\!\gamma^2(p_2\!+\!{\tan(q_{\rm f}Bs')}p_1)\Big]i\tanh(q_{\rm f}Es')\gamma^4\gamma^3\nonumber\\
&=&4\tanh(q_{\rm f}Es')\Big\{\tanh(q_{\rm f}Es)\big[m^2+(\tilde{p}_4+q_4-{i\tanh(q_{\rm f}Es)}(p_3+q_3))(\tilde{p}_4-{i\tanh(q_{\rm f}Es')}p_3)+({p}_3+q_3+{i\tanh(q_{\rm f}Es)}\nonumber\\
&&\;\;\;-(\tilde{p}_4+q_4))({p}_3+{i\tanh(q_{\rm f}Es')}\tilde{p}_4)-(p_1+q_1-{\tan(q_{\rm f}Bs)}(p_2+q_2))(p_1-{\tan(q_{\rm f}Bs')}p_2)-(p_2+q_2+{\tan(q_{\rm f}Bs)})\nonumber\\
&&\;\;\;(p_1+q_1))(p_2+{\tan(q_{\rm f}Bs')}p_1)\big]-i\big[(\tilde{p}_4+q_4-{i\tanh(q_{\rm f}Es)}(p_3+q_3))({p}_3+{i\tanh(q_{\rm f}Es')}\tilde{p}_4)-({p}_3+q_3+{i\tanh(q_{\rm f}Es)}\nonumber\\
&&\;\;\;(\tilde{p}_4+q_4))(\tilde{p}_4-{i\tanh(q_{\rm f}Es')}p_3)\big]-\tan(q_{\rm f}Bs)\tanh(q_{\rm f}Es)\big[(p_1+q_1-{\tan(q_{\rm f}Bs)}(p_2+q_2))(p_2+{\tan(q_{\rm f}Bs')}p_1)\nonumber\\
&&\;\;\;-(p_2+q_2+{\tan(q_{\rm f}Bs)}(p_1+q_1))(p_1-{\tan(q_{\rm f}Bs')}p_2)\big]\Big\}.
\end{eqnarray}
Thus, by integrating all the terms together, we find that all the cross terms among different energy-momentum components are canceled out exactly and the formula becomes simple:
\begin{eqnarray}
i\tilde{\cal D}_\sigma^{-1}(q)
&=&\frac{1}{2G}-4N_{c}\sum_{\rm f=u,d}\sum_{\tilde{p}}\int_0^\infty {ds}\int_0^\infty {ds'}\exp\Big\{-im^2s-{if_{s}[(\tilde{p}_4+q_4)^2+(p_3+q_3)^2]\over q_{\rm f}E}-{ig_{s}[(p_1+q_1)^2+(p_2+q_2)^2]\over q_{\rm f}B}\nonumber\\
&&-im^2s'-{if_{s'}(\tilde{p}_4^2+p_3^2)\over q_{\rm f}E}-{ig_{s'}(p_1^2+p_2^2)\over q_{\rm f}B}\Big\}~\Big\{m^2\big(1-g_{s}g_{s'}\big)\big(1+f_{s}f_{s'}\big)-\big[(\tilde{p}_4+q_4)\tilde{p}_4+({p}_3+q_3){p}_3\big]
\big(1-f_{s}^2\big)\nonumber\\
&&\big(1-f_{s'}^2\big)\big(1-g_{s}g_{s'}\big)-\big[({p}_2+q_2){p}_2+({p}_1+q_1){p}_1\big]\big(1+g_{s}^2\big)\big(1+g_{s'}^2\big)\big(1+f_{s}f_{s'}\big)\Big\}.
\end{eqnarray}
where we have defined $f_{s}=\tanh(q_{\rm f}Es)$ and $g_{s}=\tan(q_{\rm f}Bs)$ as in the text.\\
\indent Then the inverse propagator of $\pi_0$ mode can be obtained directly from the $\sigma$ mode, that is,
\begin{eqnarray}
i\tilde{\cal D}_0^{-1}(q)
&=&\frac{1}{2G}-N_{c}\sum_{\rm f=u,d}\sum_{\tilde{p}}\int_0^\infty {ds}\int_0^\infty {ds'}\exp\Big\{-im^2s-{i((\tilde{p}_4+q_4)^2+(p_3+q_3)^2)\over q_{\rm f}E\coth(q_{\rm f}Es)}-{i((p_1+q_1)^2+(p_2+q_2)^2)\over q_{\rm f}B\cot(q_{\rm f}Bs)}\nonumber\\
&&-im^2s'-{i(\tilde{p}_4^2+p_3^2)\over q_{\rm f}E\coth(q_{\rm f}Es')}-{i(p_1^2+p_2^2)\over q_{\rm f}B\cot(q_{\rm f}Bs')}\Big\}\;tr\Big[m-\gamma^4(\tilde{p}_4+q_4-{i\tanh(q_{\rm f}Es)}(p_3+q_3))-\gamma^3({p}_3+q_3\nonumber\\
&&+{i\tanh(q_{\rm f}Es)}(\tilde{p}_4+q_4))-\gamma^1(p_1+q_1-{\tan(q_{\rm f}Bs)}(p_2+q_2))-\gamma^2(p_2+q_2+{\tan(q_{\rm f}Bs)}(p_1+q_1))\Big]\nonumber\\
&&\Big[1+i\tan(q_{\rm f}Bs)\tanh(q_{\rm f}Es)\gamma^5+\tan(q_{\rm f}Bs)\gamma^1\gamma^2+i\tanh(q_{\rm f}Es)\gamma^4\gamma^3\Big]\Big[-m-\gamma^4(\tilde{p}_4-{i\tanh(q_{\rm f}Es')}p_3)\nonumber\\
&&-\gamma^3({p}_3+{i\tanh(q_{\rm f}Es')}\tilde{p}_4)-\gamma^1(p_1-{\tan(q_{\rm f}Bs')}p_2)-\gamma^2(p_2+{\tan(q_{\rm f}Bs')}p_1)\Big]\Big[1+i\tan(q_{\rm f}Bs')\tanh(q_{\rm f}Es')\gamma^5\nonumber\\
&&+\tan(q_{\rm f}Bs')\gamma^1\gamma^2+i\tanh(q_{\rm f}Es')\gamma^4\gamma^3\Big]\nonumber\\
&=&\frac{1}{2G}-4N_{c}\sum_{\rm f=u,d}\sum_{\tilde{p}}\int_0^\infty {ds}\int_0^\infty {ds'}\exp\Big\{-im^2s-{if_{s}[(\tilde{p}_4+q_4)^2+(p_3+q_3)^2]\over q_{\rm f}E}-{ig_{s}[(p_1+q_1)^2+(p_2+q_2)^2]\over q_{\rm f}B}\nonumber\\
&&-im^2s'-{if_{s'}(\tilde{p}_4^2+p_3^2)\over q_{\rm f}E}-{ig_{s'}(p_1^2+p_2^2)\over q_{\rm f}B}\Big\}~\Big\{-m^2\big(1-g_{s}g_{s'}\big)\big(1+f_{s}f_{s'}\big)-\big[(\tilde{p}_4+q_4)\tilde{p}_4+({p}_3+q_3){p}_3\big]
\big(1-f_{s}^2\big)\nonumber\\
&&\big(1-f_{s'}^2\big)\big(1-g_{s}g_{s'}\big)-\big[({p}_2+q_2){p}_2+({p}_1+q_1){p}_1\big]\big(1+g_{s}^2\big)\big(1+g_{s'}^2\big)
\big(1+f_{s}f_{s'}\big)\Big\}
\end{eqnarray}
with only a change of the sign of the mass term in the second brace as in the pure magnetic field case~\cite{Cao:2014uva}. Finally, the coefficients of $i\tilde{\cal D}_0^{-1}(q)$ around small momenta ${\bf q}$ at zero energy $q_4=0$ can be evaluated by taking Taylor expansions and we find
\begin{eqnarray}
\xi_{\bot}&=&-4N_{c}\sum_{\rm f=u,d}\sum_{\tilde{p}}\int_0^\infty {ds}\int_0^\infty {ds'}\exp\Big\{-im^2(s+s')-{i(f_{s}+f_{s'})\over q_{\rm f}E}
(\tilde{p}_4^2+{p}_3^2)-{i(g_{s}+g_{s'})\over q_{\rm f}B}
({p}_2^2+{p}_1^2)\Big\}\Big\{{ig_{s}g_{s'}\over
 q_{\rm f}B(g_{s}+g_{s'})}\nonumber\\
&&\Big[m^2(1-g_{s}g_{s'})(1+f_{s}f_{s'})+(\tilde{p}_4^2+{p}_3^2)
(1-f_{s}^2)(1-f_{s'}^2)(1-g_{s}g_{s'})+({p}_2^2+{p}_1^2)(1+g_{s}^2)(1+g_{s'}^2)(1+f_{s}f_{s'})\Big]\nonumber\\
&&+{g_{s}g_{s'}\over (g_{s}+g_{s'})^2}(1+g_{s}^2)(1+g_{s'}^2)(1+f_{s}f_{s'})\Big\}\nonumber\\
&=&-{N_{c}\over2\pi^{3/2}}\sum_{\rm f=u,d}\sum_{\tilde{p}_4}\int_0^\infty {ds}\int_0^\infty {ds'}\exp\Big\{-im^2(s+s')-{i(f_{s}+f_{s'})\over q_{\rm f}E}
\tilde{p}_4^2\Big\}{q_{\rm f}B\over i(g_{s}+g_{s'})}\Big({q_{\rm f}E\over i(f_{s}+f_{s'})}\Big)^{1/2}\Big\{{ig_{s}g_{s'}\over
 q_{\rm f}B(g_{s}+g_{s'})}\nonumber\\
&&\Big[m^2(1-g_{s}g_{s'})(1+f_{s}f_{s'})+(\tilde{p}_4^2+{q_{\rm f}E\over 2i(f_{s}+f_{s'})})
(1-f_{s}^2)(1-f_{s'}^2)(1-g_{s}g_{s'})\Big]+{2g_{s}g_{s'}\over (g_{s}+g_{s'})^2}(1+g_{s}^2)(1+g_{s'}^2)(1+f_{s}f_{s'})\Big\}\nonumber
\end{eqnarray}
\begin{eqnarray}
&=&-{N_{c}\over2\pi^{3/2}}\sum_{\rm f=u,d}\sum_{\tilde{p}_4}\int_0^\infty {ds}\int_0^\infty {ds'}\exp\Big\{-im^2(s+s')-{i(f_{s}+f_{s'})\over q_{\rm f}E}
\tilde{p}_4^2\Big\}{q_{\rm f}B\over i(g_{s}+g_{s'})}\Big({q_{\rm f}E\over i(f_{s}+f_{s'})}\Big)^{1/2}\Big\{{ig_{s}g_{s'}\over
 q_{\rm f}B(g_{s}+g_{s'})}\nonumber\\
&&\Big[m^2(1-g_{s}g_{s'})(1+f_{s}f_{s'})+{q_{\rm f}E\over 2i(f_{s}+f_{s'})}
(1-f_{s}^2)(1-f_{s'}^2)(1-g_{s}g_{s'})\Big]+{2g_{s}g_{s'}\over (g_{s}+g_{s'})^2}(1+g_{s}^2)(1+g_{s'}^2)(1+f_{s}f_{s'})\Big\}\nonumber\\
&&-i{N_{c}\over2\pi^{3/2}}\sum_{\rm f=u,d}\sum_{\tilde{p}_4}\int_0^\infty \!\!\!{d\exp\Big\{-{i(f_{s}+f_{s'})\over q_{\rm f}E}
\tilde{p}_4^2\Big\}}\int_0^\infty\!\!\! {ds'}e^{-im^2(s+s')}{q_{\rm f}B\over i(g_{s}+g_{s'})}\Big({q_{\rm f}E\over i(f_{s}+f_{s'})}\Big)^{1/2}{ig_{s}g_{s'}\over
 q_{\rm f}B(g_{s}+g_{s'})}(1-f_{s'}^2)(1-g_{s}g_{s'})\nonumber\\
&=&-{N_{c}\over4\pi^{2}}\sum_{\rm f=u,d}\int_0^\infty\!\!\! {ds}\int_0^\infty\!\!\! {ds'}e^{-im^2(s+s')}{q_{\rm f}B\over i(g_{s}\!+\!g_{s'})}
{q_{\rm f}E\over i(f_{s}\!+\!f_{s'})}\vartheta_3\left({\pi\over 2}\!+\!i{\mu\over 2T},e^{-|i{q_{\rm f}E\over 4(f_s+f_{s'})T^2}|}\right)\Big\{{ig_{s}g_{s'}\over
 q_{\rm f}B(g_{s}\!+\!g_{s'})}\Big[m^2f_{s'}(f_{s}\!+\!f_{s'})(1\!-\!g_{s}g_{s'})\nonumber\\
&&-iq_{\rm f}B\Big({g_{s'}-g_{s}\over g_{s}(g_{s}+g_{s'})}
-{g_{s'}\over (1-g_{s}g_{s'})}\Big)
(1-f_{s'}^2)(1+g_{s}^2)(1-g_{s}g_{s'})\Big]+{2g_{s}g_{s'}\over (g_{s}+g_{s'})^2}(1+g_{s}^2)(1+g_{s'}^2)(1+f_{s}f_{s'})\Big\},
\end{eqnarray}
\begin{eqnarray}
\xi_{3}&=&-4N_{c}\sum_{\rm f=u,d}\sum_{\tilde{p}}\int_0^\infty {ds}\int_0^\infty {ds'}\exp\Big\{-im^2(s+s')-{i(f_{s}+f_{s'})\over q_{\rm f}E}
(\tilde{p}_4^2+{p}_3^2)-{i(g_{s}+g_{s'})\over q_{\rm f}B}
({p}_2^2+{p}_1^2)\Big\}\Big\{{if_{s}f_{s'}\over
 q_{\rm f}E(f_{s}+f_{s'})}\nonumber\\
&&\Big[m^2(1-g_{s}g_{s'})(1+f_{s}f_{s'})+(\tilde{p}_4^2+{p}_3^2)
(1-f_{s}^2)(1-f_{s'}^2)(1-g_{s}g_{s'})+({p}_2^2+{p}_1^2)(1+g_{s}^2)(1+g_{s'}^2)(1+f_{s}f_{s'})\Big]\nonumber\\
&&+{f_{s}f_{s'}\over (f_{s}+f_{s'})^2}(1-f_{s}^2)(1-f_{s'}^2)(1-g_{s}g_{s'})\Big\}\nonumber\\
&=&-{N_{c}\over2\pi^{3/2}}\sum_{\rm f=u,d}\sum_{\tilde{p}_4}\int_0^\infty {ds}\int_0^\infty {ds'}\exp\Big\{-im^2(s+s')-{i(f_{s}+f_{s'})\over q_{\rm f}E}
\tilde{p}_4^2\Big\}{q_{\rm f}B\over i(g_{s}+g_{s'})}\Big({q_{\rm f}E\over i(f_{s}+f_{s'})}\Big)^{1/2}\Big\{{if_{s}f_{s'}\over
 q_{\rm f}E(f_{s}+f_{s'})}\nonumber\\
&&\Big[m^2(1-g_{s}g_{s'})(1+f_{s}f_{s'})+\tilde{p}_4^2
(1-f_{s}^2)(1-f_{s'}^2)(1-g_{s}g_{s'})+{q_{\rm f}B\over i(g_{s}+g_{s'})}(1+g_{s}^2)(1+g_{s'}^2)(1+f_{s}f_{s'})\Big]\nonumber\\
&&+{3f_{s}f_{s'}\over 2(f_{s}+f_{s'})^2}(1-f_{s}^2)(1-f_{s'}^2)(1-g_{s}g_{s'})\Big\}\nonumber\\
&=&-{N_{c}\over2\pi^{3/2}}\sum_{\rm f=u,d}\sum_{\tilde{p}_4}\int_0^\infty {ds}\int_0^\infty {ds'}\exp\Big\{-im^2(s+s')-{i(f_{s}+f_{s'})\over q_{\rm f}E}
\tilde{p}_4^2\Big\}{q_{\rm f}B\over i(g_{s}+g_{s'})}\Big({q_{\rm f}E\over i(f_{s}+f_{s'})}\Big)^{1/2}\Big\{{if_{s}f_{s'}\over
 q_{\rm f}E(f_{s}+f_{s'})}\nonumber\\
&&\Big[m^2(1-g_{s}g_{s'})(1+f_{s}f_{s'})
+{q_{\rm f}B\over i(g_{s}+g_{s'})}(1+g_{s}^2)(1+g_{s'}^2)(1+f_{s}f_{s'})\Big]+{3f_{s}f_{s'}\over 2(f_{s}+f_{s'})^2}(1-f_{s}^2)(1-f_{s'}^2)(1-g_{s}g_{s'})\Big\}\nonumber\\
&&-i{N_{c}\over2\pi^{3/2}}\sum_{\rm f=u,d}\sum_{\tilde{p}_4}\int_0^\infty\!\!\! {d\exp\Big\{-{i(f_{s}+f_{s'})\over q_{\rm f}E}
\tilde{p}_4^2\Big\}}\int_0^\infty\!\!\! {ds'}e^{-im^2(s+s')}{q_{\rm f}B\over i(g_{s}+g_{s'})}\Big({q_{\rm f}E\over i(f_{s}+f_{s'})}\Big)^{1/2}\!\!\!{if_{s}f_{s'}\over
 q_{\rm f}E(f_{s}+f_{s'})}(1-f_{s'}^2)(1-g_{s}g_{s'})\nonumber\\
&=&-{N_{c}\over4\pi^{2}}\sum_{\rm f=u,d}\int_0^\infty {ds}\int_0^\infty {ds'}e^{-im^2(s+s')}{q_{\rm f}B\over i(g_{s}+g_{s'})}
{q_{\rm f}E\over i(f_{s}+f_{s'})}\vartheta_3\left({\pi\over 2}+i{\mu\over 2T},e^{-|i{q_{\rm f}E\over 4(f_s+f_{s'})T^2}|}\right)\Big\{{if_{s}f_{s'}\over
 q_{\rm f}E(f_{s}+f_{s'})}\Big[m^2f_{s'}(f_{s}+f_{s'})\nonumber\\
&&(1-g_{s}g_{s'})-{iq_{\rm f}B\over (g_{s}+g_{s'})}(1+g_{s}^2)(1+g_{s'}^2)(1+f_{s}f_{s'})+iq_{\rm f}B\Big({1\over g_{s}+g_{s'}}
+{g_{s'}\over 1-g_{s}g_{s'}}\Big)
(1-f_{s'}^2)(1+g_{s}^2)(1-g_{s}g_{s'})\Big]\nonumber\\
&&+{f_{s'}\over f_{s}+f_{s'}}(1-f_{s}^2)(1-f_{s'}^2)(1-g_{s}g_{s'})\Big\}.
\end{eqnarray}
In the last steps of derivations, partial integrations for the proper time $s$ are performed firstly to remove the initial $\tilde{p}_4^2$ term in the second brace. Then the summations over the Matsubara frequency denoted by $\tilde{p}_4$ are completed which give rise to the third Jacobi theta function $\vartheta_3(z,q)$.
\end{widetext}

\end{document}